%% file: main.tex
\documentclass[conference]{IEEEtran}
\pdfoutput=1
\usepackage{balance}  
\usepackage{graphicx} 
\usepackage{multirow}
\usepackage{array,multirow}
\usepackage{hhline}
\usepackage{times}    
\usepackage{url}   
\usepackage{cite} 
\usepackage{tikz}
\usepackage{verbatim}
\usepackage{rotating}

\usepackage{lipsum}
\usepackage{chngcntr}

\usepackage{multirow}
\usepackage{amsmath}
\usepackage[caption=false]{subfig}

\definecolor{Gray}{gray}{0.8}
\usepackage{array}
\usepackage{fancyhdr}
\usepackage[T1]{fontenc}
\usepackage{colortbl, booktabs}
\begin{document}

\title{Choosing Requirements for Experimentation with User 
Interfaces of Requirements Modeling Tools
 }

\author{
\IEEEauthorblockN{Parisa Ghazi \IEEEauthorrefmark{1}, Zahra Shakeri Hossein Abad\IEEEauthorrefmark{2}, Martin Glinz\IEEEauthorrefmark{1}}
 \IEEEauthorblockA{\IEEEauthorrefmark{1}Department of Informatics, University of Zurich, Switzerland
    \\\{ghazi, glinz\}@ifi.uzh.ch}
    \IEEEauthorblockA{\IEEEauthorrefmark{2}SEDS Lab, Department of Computer Science, University of Calgary, Canada
    \\zshakeri@ucalgary.ca}
}

\maketitle

\begin{abstract}
When designing a new presentation front-end called FlexiView for requirements modeling tools, we encountered a general problem: designing such an interface requires a lot of experimentation which is costly when the code of the tool needs to be adapted for every experiment. On the other hand, when using simplified user interface (UI) tools, the results are difficult to generalize. To improve this situation, we are developing an UI experimentation tool which is based on so-called ImitGraphs. ImitGraphs can act as a simple, but accurate substitute for a modeling tool. In this paper, we define requirements for such an UI experimentation tool based on an analysis of the features of existing requirements modeling tools.

\end{abstract}


\begin{IEEEkeywords}
Graphical Models, Requirements Engineering, Modeling Tools, User Interface
\end{IEEEkeywords}


\IEEEpeerreviewmaketitle

\input{sections/01-introduction.tex}

\input{sections/02-ToolsComparison.tex}

\input{sections/03-theFlexiviewTool.tex}

\input{sections/04-Scenario.tex}

\input{sections/05-conclusions-and-futurework.tex}

\bibliographystyle{IEEEtran}

\bibliography{main}
\end{document}

%% file: sections/01-introduction.tex
\section{Introduction}
\label{introduction}
Requirements engineers spend a lot of their time working with modeling tools. Thus, the usability of their modeling tools affects their productivity\cite{de2012requirements}. However, the User Interface (UI) of this type of tools has not changed for a long time despite the challenges that exist in working with artifacts \cite{RE16}. Information presentation is one of the aspects of the modeling tools that can be improved. We are developing a new tool front-end called FlexiView~\cite{REFSQ2015} for using the screen space efficiently by presenting information in heterogeneous levels of detail. Like every other new feature, it should go through multiple cycles of usability experimentation and optimization in order to mature.

The high cost of usability experiments at the early stages of software development is one of the reasons that the improvement of the UIs of modeling tools are neglected. We have proposed ImitGraphs~\cite{MiSE17} 
to lower the cost of usability experiments. ImitGraphs are an extended version of Graphs that can substitute Requirements Engineering (RE) graphical models (e.g., diagrams such as activity diagrams and sequence diagrams) in usability experiments. The simplicity of ImitGraphs enables usability testers to quickly develop experimental tools instead of using the modeling tools as a testing platform. 
We intend to design an experimental tool based on ImitGraphs for testing and optimizing FlexiView. 
Since FlexiView will be integrated into modeling tools, the experimental tool that we design should have features similar to the features of existing modeling tools. 
Our goal in this paper is to study the basic features of the existing modeling tools and define the requirements of a suitable experimental tool by including the most frequent features. 

To achieve this goal, we conducted a market study in which we analyzed the UI features of a group of modeling tools. Then, we selected the features with the highest frequency as the UI requirements of the tool that we need for experimenting on Flexiview. 
Our contributions are (i)~a list of basic essential manipulation actions in modeling tools, (ii)~different methods of performing those actions with their frequencies, and (iii)~the UI requirements of an experimental tool for testing UI features.


%% file: sections/02-ToolsComparison.tex
\section{Approach}
\label{sec:Req}
We performed our study of basic features of existing modeling tools in three steps: selecting tools, defining basic manipulation actions and finding the most frequent method for each action. In {\bf step 1}, we defined the criteria of selecting tools for our study. Since FlexiView is designed to be used in touch screen modeling tools, we searched Google Play Store with the keywords ``UML'' and ``diagram'' and picked apps with an average score of 3.6 and above. We ended up with the following list of ten modeling tools: Flowdia Lite~(T1), Draw Express Lite~(T2), Flexi{S}ketch~(T3),	Droiddia~(T4), Grapholite~(T5), Draw.io~(T6), Lekh Diagram~(T7), Diagrid~(T8), ClickCharts Free~(T9), and NodeScape~(T10). In {\bf step 2}, we defined a set of basic manipulation actions that are essential for a modeling tool by watching modeling tutorials on YouTube. The videos were not related to the tools under study and were intended for beginners. We extracted the following list of eleven basic essential actions from the videos: creating a new object, opening a context menu, scrolling, deleting an existing object or an existing connection, selecting multiple objects, duplicating an object, changing the color of an object, changing the style of a connection, moving an object, connecting objects, and adding/changing the text of an object or a connection.
%
In {\bf step 3}, we inspected the tools and identified the methods by which the actions can be performed.

{
\begin{table}
\setlength\tabcolsep{1.5pt}
\tiny
\centering
\caption{Basic actions and how they can be performed in tools}
\label{tab:toolcomparison}

\begin{tabular}{|>{\centering}m{.36\linewidth}|c|>{\centering}m{.01\linewidth}|c|}
\hline
{\bf  Action}  & {\bf Method} & {\bf \#} & {\bf Tool}  \\\hline
 
    \multirow{5}{*}{\parbox{2cm}{\bf\centering Create an object}}&
    \cellcolor{Gray} Drag and drop from the menu&\cellcolor{Gray} 5&T1-2, T6, T9-10\\ 
      &\cellcolor{Gray} Select from the menu &\cellcolor{Gray}{5}&T1-2, T5-7\\  
      &Free draw &2&T3, T7\\   
     &Long touch then select from the pop-up menu&1&T4\\   
     &Select from the menu then single touch &3& T3, T8-9\\
     \hline
     \hline 
\multirow{3}{*}{\parbox{1.5cm}{\bf\centering Open the Context menu}}
    &\cellcolor{Gray}Same time as selecting the object&\cellcolor{Gray}8& T1-5, T7-8, T10\\
   &Single touch on the selected object  &1&T6\\
  &Not available &1&T9\\ 
      
     \hline
     \hline 
     \multirow{2}{*}{\bf Scrolling}
  &\cellcolor{Gray} One finger&\cellcolor{Gray}7& T1, T4-10\\ 
     &Two fingers&3&T2-3, T6\\  
    
     \hline
     \hline 
   \multirow{2}{*}{\bf Delete object/connection}&
    \cellcolor{Gray} Select from the context menu &\cellcolor{Gray}{8}& T1, T3-8, T10\\  
      &Select from the menu &2&T2, T9\\   
     \hline
     \hline 
     
  \multirow{3}{*}{\parbox{1.8cm}{\bf\centering Select multiple objects: \it{Step~1-initiate selection}} } 
        &Long touch on the canvas & 3&T5-6, T9\\ 
      &\cellcolor{Gray}Select from the menu & \cellcolor{Gray}5 &T1-2, T7, T9-10\\ 
      &Not available &3&T3-4, T8\\ \hline
      \multirow{2}{*}{\parbox{1.8cm}{\bf\centering Select multiple objects: \it{Step~2-indicate objects }} }
     &Free hand & 1&T2\\
      &\cellcolor{Gray}Rectangle &\cellcolor{Gray}{6}& T1, T5-7, T9-10 \\   
     \hline
     \hline

   \multirow{3}{*}{\bf Duplicate an object}&  
      \cellcolor{Gray} Select from the context menu & \cellcolor{Gray}6 &T1, T4-7, T10\\ 
      &Gesture command &1&T2\\   
     &Not available & 3&T3, T8-9\\
     \hline
     \hline 
   \multirow{4}{*}{\bf Edit object color}&
   \cellcolor{Gray}Select from the context menu&\cellcolor{Gray}5& T1, T4, T7-8, T10 \\ 

     & Edit directly in the side menu &{4}&T2, T5-6, T9\\  
      & Double click on the object &1&T8\\
      &Not available &1&T3\\
     \hline
     \hline     
\multirow{3}{*}{{\parbox{1.5cm}{\bf\centering Change a connection's style}}}&
   \cellcolor{Gray}Select from the context menu&\cellcolor{Gray}6& T1,T3-4,T7-8,T10 \\ 
     & Edit directly in the side menu &4&T2, T5-6, T9\\
      & Double click on the connection &1&T8\\
    
     \hline
     \hline 
   
\multirow{3}{*}{\bf Move an object: \it{Step~1-initiate move}}
&\cellcolor{Gray}  Move selected&\cellcolor{Gray}{7}&T2, T3, T5-8, T10\\ 
   
   & Move unselected & 3& T1, T6, T9 \\ 

     & Long touch &1&T4\\  \hline  
     \multirow{2}{*}{\bf Move an object: \it{Step~2-move}}
     & Dragging the handle &2&T2, T10\\
    
     &\cellcolor{Gray}Dragging the object&\cellcolor{Gray}{8}&T1, T3-9\\ 
     
     \hline
     \hline
     
\multirow{5}{*}{\bf Connect two objects}&

Drag the handle &2&T5-6\\ 
   
   & Select from the menu & 1& T6   \\ 
& Select from the menu then draw a free hand conn & 2& T9-10  \\ 
     & \cellcolor{Gray}Select from context menu then select second obj &\cellcolor{Gray}3&T1, T4, T8\\  
     & \cellcolor{Gray}Draw free hand connection &\cellcolor{Gray}3&T2-3, T7\\
     \hline
     \hline 
\multirow{2}{*}{\bf Change the text}&
\cellcolor{Gray}Double click on the object and connection &
\cellcolor{Gray}7&T2, T4-6, T8-10\\ 
   & Select from the context menu & 4& T1, T3, T7, T8 \\ 
     \hline
\end{tabular}
\end{table}
}

%% file: sections/03-theFlexiviewTool.tex
\section{Results}
\label{theFlexiviewTool }
The result of our observations is presented in Table~\ref{tab:toolcomparison}. The columns contain essential actions, the methods of performing those actions, the number of the tools that employ those methods, and the corresponding tools.
For example, creating an object can be done differently, e.g., by dragging the object from the menu and dropping it on the canvas, by selecting the object from the menu, by freehand drawing the object, by long touching a location on the canvas and selecting the object from the pop-up menu, or by selecting the object from the menu and then touching a location on the canvas. 

Some actions can be performed in more than one way in some tools. This is the reason why, for some actions, the sum of the frequencies is more than the number of the studied tools. Some actions can be performed in separate independent steps. For example, for selecting multiple objects, first, the user initiates the selection, then, indicates the objects. The first step can be done by a long touch, or by selecting the corresponding icon from the menu. The second step can be done by drawing a freehand lasso around the objects, or by drawing a rectangle around the objects. Considering such steps separately allowed us to find the most frequent methods more accurately than by analyzing actions only, since an action might be rather infrequent while one of its constituent steps occurs frequently.

%% file: sections/04-Scenario.tex
\section{Choosing the Requirements}
\label{sec:sc}
Based on the frequency of the methods, we chose the requirements for our experimental tool.
During this process, we encountered two special cases. First, when two methods conflicted, and second when there was a tie. In the case of a conflict, we chose the combination of methods with a higher overall frequency. For example, we cannot have scrolling with one finger and connecting objects by freehand drawing at the same time.
We had to choose between (i)~one-finger scrolling and using the context menu for connecting objects, or (ii)~two-finger scrolling and connecting objects by freehand drawing.
In this case, we chose the first combination based on the higher overall frequency.

In case of a tie, if implementing both of the options was possible, we chose both.
For example, creating an object by drag-and-drop or by selection from the menu could co-exist in a tool. Therefore, we chose both of them.
If methods with equal frequencies could not co-exist in a tool, we used the score of the tools to break the tie.
Finally, our study resulted in the following requirements for our experimental tool.

{\it{
1.~The tool should allow users to create an object by drag-and-dropping it from the menu onto the canvas and also by selecting the icon of an object from the menu. 2.~A context menu should appear when an object is selected. 3.~The user should be able to scroll using one finger. 4.~An object can be deleted by selecting the corresponding command from the context menu. 5.~To select multiple objects, the user should first select the corresponding command from the menu and then draw a rectangle around the desired objects. 6.~In order to duplicate an object, the corresponding command should be selected from the context menu. 
7.~The user should be able to change the color of the objects after selecting the corresponding command from the context menu.
8.~The user should be able to change the color and type of the connections after selecting the corresponding command from the context menu. 9.~The user should be able to move a selected object by dragging. 10.~In order to connect two objects, first the corresponding command should be selected from the context menu and then the second object should be selected. 11.~The user should be able to change the text of the objects and connections after double-tapping on them.

}}

%% file: sections/05-conclusions-and-futurework.tex
\section{Conclusion and Future Work}
\label{sec:CF}

When a feature such as FlexiView will be eventually integrated into other modeling tools, the generalizability of the usability experiments is important.
Therefore, the UI features of the experimental tool should be as similar as possible to the features of the target modeling tools. In order to design such a tool, we studied 
available modeling tools and extracted the most frequent methods of performing essential actions in those tools. Based on the results, we defined the UI requirements of an ImitGraphs-based, experimental tool that can be used for experimentation with the UI of RE tools.  

This work will be continued by actually implementing a tool based on the defined requirements and conducting usability experiments with the UI of FlexiView.

%% file: main.bbl
\begin{thebibliography}{1}
\providecommand{\url}[1]{#1}
\csname url@samestyle\endcsname
\providecommand{\newblock}{\relax}
\providecommand{\bibinfo}[2]{#2}
\providecommand{\BIBentrySTDinterwordspacing}{\spaceskip=0pt\relax}
\providecommand{\BIBentryALTinterwordstretchfactor}{4}
\providecommand{\BIBentryALTinterwordspacing}{\spaceskip=\fontdimen2\font plus
\BIBentryALTinterwordstretchfactor\fontdimen3\font minus
  \fontdimen4\font\relax}
\providecommand{\BIBforeignlanguage}[2]{{%
\expandafter\ifx\csname l@#1\endcsname\relax
\typeout{** WARNING: IEEEtran.bst: No hyphenation pattern has been}%
\typeout{** loaded for the language `#1'. Using the pattern for}%
\typeout{** the default language instead.}%
\else
\language=\csname l@#1\endcsname
\fi
#2}}
\providecommand{\BIBdecl}{\relax}
\BIBdecl

\bibitem{de2012requirements}
J.~M.~C. De~Gea, J.~Nicol{\'a}s, J.~L.~F. Alem{\'a}n, A.~Toval, C.~Ebert, and
  A.~Vizca{\'\i}no, ``Requirements engineering tools: Capabilities, survey and
  assessment,'' \emph{Information and Software Technology}, vol.~54, no.~10,
  pp. 1142--1157, 2012.

\bibitem{RE16}
P.~Ghazi and M.~Glinz, ``An exploratory study on user interaction challenges
  when handling interconnected requirements artifacts of various sizes,'' in
  \emph{24th {IEEE} International Requirements Engineering Conference (RE
  '16)}.\hskip 1em plus 0.5em minus 0.4em\relax IEEE, 2016, pp. 76--85.

\bibitem{REFSQ2015}
P.~Ghazi, N.~Seyff, and M.~Glinz, ``Flexi{V}iew: {A} magnet-based approach for
  visualizing requirements artifacts,'' in \emph{21st International Working
  Conference on Requirements Engineering: Foundation for Software Quality
  (REFSQ '15)}.\hskip 1em plus 0.5em minus 0.4em\relax Springer, 2015, pp.
  262--269.

\bibitem{MiSE17}
P.~Ghazi and M.~Glinz, ``Imit{G}raphs: Towards faster usability tests of
  graphical model manipulation techniques,'' in \emph{9th International
  Workshop on Modeling in Software Engineering (MiSE@ICSE2017)}, 2015.

\end{thebibliography}
